\documentclass[a4paper]{spie}  %>>> use this instead for A4 paper

\usepackage{amsmath,amsfonts,amssymb}
\usepackage{graphicx}
\usepackage[colorlinks=true, allcolors=blue]{hyperref}
\usepackage{bm}

\title{Ultrafast spin current generation in ferromagnetic thin films}

\author[]{Giovanni Manfredi}
\author[]{J\'er\^ome Hurst}
\author[]{Paul-Antoine Hervieux}

\affil[]{Universit\'e de Strasbourg, CNRS, Institut de Physique et Chimie des Mat\'eriaux de Strasbourg, UMR 7504, F-67000 Strasbourg, France}
\authorinfo{Further author information: (Send correspondence to Giovanni Manfredi)\\Giovanni Manfredi: E-mail: giovanni.manfredi@ipcms.unistra.fr, Telephone: +33 (0)3 88 10 72 14}

\begin{document}
\maketitle

\begin{abstract}
Spin currents have been shown to play
a key role in the ultrafast laser-driven demagnetization
process in ferromagnetic thin films. Here, we show that an oscillating spin current can be generated in the film via the application of a femtosecond laser pulse in
the visible range. In particular, we prove that the spin current is due to ballistic electrons traveling back and forth in the film. The dependence of the current intensity on the driving electric field is also investigated.
\end{abstract}


\section{INTRODUCTION}
\label{sec:intro}

The interaction of a femtosecond electromagnetic pulse with the electronic spins in a ferromagnetic nanostructure has been the object of intense investigations, both theoretical  and experimental, over the past two decades. Experimentally, the main effect that has been observed -- though not yet fully elucidated -- is the quick loss of magnetization following the excitation by the laser pulse \cite{Beaurepaire1996}. Several mechanisms have been proposed for the modification of the magnetic order of nanostructures subject to an ultrafast external field, ranging from the spin-orbit coupling  to spin-lattice interactions  and superdiffusive spin currents \cite{Battiato2010}. However, the evidence is still inconclusive, perhaps pointing to a multiplicity of underlying mechanisms that are material-dependent.

From a fundamental point of view, the electromagnetic field associated with a femtosecond laser pulse is often strong enough to significantly perturb the electronic charges and spins in condensed-matter systems, so that relativistic effects such as the spin-orbit coupling may become important. Given the intensity of the fields involved, nonlinear effects are also expected to play a considerable role.

Computational work aimed at understanding the ultrafast magnetization dynamics has mainly developed along two lines: (i) Earlier works employed either macroscopic approaches based on the three-temperature model \cite{Beaurepaire1996} or phenomenological Boltzmann-type equations within the framework of Fermi-liquid theory \cite{Guillon2003,Aeschlimann2000}; (ii) more recently, Gross and co-workers  developed a first-principles computational code based on time-dependent density functional theory (TDDFT) that includes the spin degrees of freedom as well as the spin-orbit coupling \cite{Krieger2015}. The latter approach is particularly attractive because it can potentially incorporate all relevant microscopic mechanisms, but has a very high computational cost and, like most other wavefunction-based methods, cannot deal with non-unitary (dissipative) effects.

In order to tackle this complex problem, we have recently developed an approach based on the phase-space representation of quantum mechanics due to Wigner \cite{Manfredi2005film,Jasiak2010}. In this formulation, the state of a quantum system is represented by a function of the phase space variables (position and momentum) plus time. It can deal on an equal footing with both pure and mixed quantum states and is equivalent to the more standard approach based on the density matrix. Further, the Wigner equation of motion bears a strong similarity with the corresponding Vlasov  equation, which constitutes its classical limit. More recently, this approach was generalized to include the spin dynamics \cite{Zamanian2010nj,Hurst2014,Hurst2017}.

In this work, we focus our attention on the generation and control of ultrafast spin currents, which have been shown to play a key role in the demagnetization process \cite{Schellekens2014}, with important applications to spintronic devices.
In particular, we show that a relatively low-amplitude laser pulse can be used to trigger a significant spin current in a nickel film \cite{Hurst2018}. Such spin current results from the ballistic motion of polarized electrons bouncing back and forth on the film surfaces, and may be observed through its emitted magnetic dipole radiation.

\section{PHASE-SPACE MODEL}
\label{sec:model}
Our purpose is to model simultaneously the charge and spin dynamics in a ferromagnetic nickel film. Concerning the spin dynamics, the main idea of our approach is to make a distinction between {\em itinerant} and {\em localized} magnetism. The former will be described by a kinetic spin-Vlasov equation, whereas the latter will be modeled through a Landau-Lifschitz-Gilbert (LLG) equation -- the two descriptions being coupled by the electron-ion magnetic exchange interactions \cite{Hurst2018}.

The electronic structure of nickel is: [Ni] = [Ar] $3d^{8}$ $4s^{2}$. Thus, there are ten valence electrons, eight of which are in the $3d$ band and the remaining two in the $4s$ band. The delocalized $4s$ electrons are described by our set of spin-Vlasov equations and thus possess both an orbital and a spin angular momentum.
In contrast, the $3d^{8}$ electrons (together with their nuclei) are assumed to form a localized and immobile ionic distribution, whose spin orientation can precess on the unit sphere \cite{Stohr2006}.

We consider a nickel film of thickness $L \approx 5 \rm nm$ and adopt a jellium model for the ionic density:
\begin{equation}
n_{i}(x) = \frac{\overline n_i}{1+\exp \left(\frac{|x| -L/2}{\sigma}\right)} ,
\label{eq:ionic density}
\end{equation}
where $\sigma \approx 0.1 \rm nm $ is a soft-edge parameter, $\overline n_i=n_0/2$ is the bulk ion density, and $n_0$ is the density of the itinerant $4s$ electrons (the factor 1/2 comes from the fact that the ions are doubly charged).
This one-dimensional (1D) Ansatz holds if the film extensions in the directions parallel to its surfaces are large compared to $L$. In such case, it is appropriate to use a 1D model, where only the coordinate $x$, normal to the film surfaces, plays a role. The relevant phase space is thus 2D: $(x,v)$.

\subsection{Itinerant magnetism (electrons)}

For the itinerant electron dynamics, we use a Wigner function approach \cite{Hurst2014,Hurst2017}. For a particle with spin, the Wigner pseudo-distribution function is actually a $2\times 2$ matrix:
\begin{equation}
\mathcal{F} (x,v,t)=
\begin{pmatrix}
f^{\uparrow  \uparrow} & f^{\uparrow  \downarrow}  \\
 f^{\downarrow  \uparrow} &  f^{\downarrow  \downarrow}
\end{pmatrix}.
\end{equation}
Using the Pauli basis, the four components of such Wigner matrix can be written in the form of a scalar phase-space distribution $f_0(x,v,t)$ and a vector distribution $f_k(x,v,t)$, with $k=\{x,y,z\}$, defined as:
\begin{equation}
f _{0} = \textrm{tr} [ \mathcal{F}]  = f ^{\uparrow \uparrow} + f ^{\downarrow \downarrow},  ~~~~~~
\bm{f}   = \frac{\hbar}{2} \textrm{tr} \left( \mathcal{F} \bm \sigma \right) ,
\end{equation}
where $\bm \sigma$ are the Pauli matrices.
In practice, $f_0$ is the analog of the standard Wigner function for particles without spin yielding the probability to find an electron at a certain location in the phase space, while $f_k$ is related to the spin and represents the probability to find an electron whose spin is oriented along the direction $k$.
In the semiclassical limit, the 1D Wigner functions obey the following transport equations:
\begin{align}
&\frac{\partial f_{0}}{\partial t}
+
v  \frac{\partial f_{0}}{\partial x}
-
\frac{1}{m} \frac{\partial V}{\partial x} \frac{\partial f_0}{\partial v} - \sum_{j=\{x,y,z\}} \frac{\mu_B}{m} \frac{\partial B_j}{\partial x} \frac{\partial f_j}{\partial v}= 0, \label{eq:spin-vlasov1}
\\
&
\frac{\partial f_{j}}{\partial t}
+
v \frac{\partial f_{j}}{\partial x}
-
\frac{1}{m} \frac{\partial V}{\partial x}\frac{\partial f_{j}}{\partial v} -
\frac{\mu_{B}}{m} \frac{\partial B_{j}}{\partial x} \frac{\partial f_{0}}{\partial v} -
\frac{e}{m}  \left[\bm{B}  \times \bm{f}\right]_{j} = 0,
\label{eq:spin-vlasov2}
\end{align}
where $\mu_{B}=e\hbar/(2m)$ is Bohr's magneton, and the scalar potential $V$ and the magnetic field $\bm{B}$ are defined as $V = -eV_H + V_{xc} + V_{ext}$ and $\bm{B} =  \bm{B}_{sd} + \bm{B}_{xc} + \bm{B}_{ext}$. $V_H$ is the Hartree potential, which satisfies the Poisson equation:
\begin{equation}
\epsilon_0 \,\partial_{x}^{2} V_{H} =  e(n_e -2n_{i})\, .
\end{equation}
$\bm{B}_{sd} = - K n_{i} \bm{S}^{i}/ 2 \mu_{B}$ is the local magnetic field due the  electron-ion exchange interactions, $V_{xc}$ and $\bm{B}_{xc}$ represent exchange-correlation effects in the local spin density approximation \cite{Perdew1981}, and $V_{ext}$ and $\bm{B}_{ext}$ denote the external fields.
In addition, one might also include the self-consistent magnetic field $\bm{B}_{A}$ obtained from Amp\`ere's equation:
\begin{equation}
\nabla \times \bm{B}_{A} = -\mu_0 e\int f_0 v dv + \mu_0 \mu_B \nabla \times \int \bm{f} dv .
\label{eq:ampere}
\end{equation}
However, $\bm{B}_{A}$ is usually very small and Amp\`ere's equation is thus discarded in most of our simulations.

The above Vlasov and Maxwell equations \eqref{eq:spin-vlasov1}-\eqref{eq:ampere} govern the dynamics of the itinerant magnetism.

\subsection{Localized magnetism (ions)}
In contrast, the localized magnetism is described  by the precession motion of the spin vector $\bm{S}^{i}(x,t)$, which obeys the LLG equation:
\begin{align}
\frac{\partial \bm{S}^{i}}{\partial t} = \frac{a^{2}J}{\hbar} \bm{S}^{i} \times  \frac{\partial^{2} \bm{S}^{i}}{\partial x^{2}} - \gamma \bm{S}^{i} \times  \bm{B}_{eff},
\label{eq:LLG}
\end{align}
where $\gamma = 2\mu_{B}/ \hbar$ is the gyromagnetic ratio, $a= 2r_{s}$ is the interatomic distance,  $J$ is the ion-ion magnetic exchange constant, $\bm{B}_{eff} = \bm{B}_{ext} + K n_0 \bm{M}^{e} / 4\mu_{B}^{2}$ is an effective magnetic field, and $\bm M ^{e}= -\frac{\mu_B}{n_0}\int \bm f dv$ is the magnetization density of the itinerant electrons.
Note that the equations for the itinerant magnetism [Eqs. \eqref{eq:spin-vlasov1}-\eqref{eq:spin-vlasov2}] are coupled to the LLG equation for the localized spins [Eq. \eqref{eq:LLG}] through the coupling constant $K$, which represents the ion-electron magnetic exchange constant.

The above set of equations \eqref{eq:spin-vlasov1}-\eqref{eq:LLG} constitute a relatively simple, but sufficiently rich, model to describe the ultrafast charge and spin dynamics in ferromagnetic nanostructures. It treats the itinerant electrons dynamics in a classical fashion, but retains the quantum-mechanical character of the spin variable (through the Wigner matrix formalism). This model will be used in the next sections to study the generation of spin currents through the excitation of the charge carriers by a short laser pulse.

In the rest of this work, we will frequently use dimensionless units, whereby energy is  normalized to the Fermi energy $E_{F}$, time to the plasmon period $T_p=2\pi\omega_p^{-1}$, where $\omega_{p} = \sqrt{e^{2} n_{0} / (m \epsilon_{0})}$, velocities to the Fermi velocity $v_{F} = \sqrt{2 E_{F}/m}$, lengths to $L_{F} = v_{F}/\omega_{p}$, and densities to $\overline n_{i} = 3/(4\pi r_{s}^{3})$, where $r_{s}$ is the Wigner-Seitz radius.
In the case of a nickel film ($r_{s}=2.6~\rm a.u.$), one has: $\overline n_{i} =n_0/2= \rm 91.8 ~nm^{-3}$,  $T_p= \rm  0.26~fs$, $v_{F} =  2.03~ \rm nm/fs$, $ L_{F} = 0.084~ \rm nm$, and $E_{F} = 11.76~\rm eV$.

\section{GROUND STATE PROPERTIES}
\label{sec:groundstate}

For spinless particles, the classical Vlasov equation for the electron distribution function $f$ is given by:
\begin{equation}
\frac{\partial f}{\partial t} + \bm{v} \cdot \bm{\nabla}_{\bm{r}}f
- \frac{e}{ m } \left(\bm{E} + \bm{v} \times \bm{B}  \right) \cdot \bm{\nabla}_{\bm{v}}  f  =  0.
\end{equation}
It can be shown that any stationary solution of the above Vlasov equation can be written as: $f_{stat} = f(H)$, where $H = \frac{m}{2}|\bm{v}|^{2} - e V_H$ is the Hamiltonian.
To satisfy the Pauli exclusion principle, the stationary state should be a Fermi-Dirac distribution function:
\begin{equation}
f_{stat} = n_{0} \mathcal{F_{D}}\left( H \right) = n_{0} \mathcal{F_{D}}\left( \frac{m}{2} |\bm{v}|^{2} - e V_H\right),
\label{eq:fstat}
\end{equation}
where
\begin{equation}
\mathcal{F_{D}} (x) =  \left[ 1 +\exp \left( \frac{x - \mu}{k_{B}T} \right) \right]^{-1},
\label{3d fermi dirac}
\end{equation}
with $\mu$ being the chemical potential. As $f_{stat}$ depends on the self-consistent electric potential $V_H$, such stationary solution is purely formal. An explicit solution can be obtained by plugging Eq. \eqref{eq:fstat} into Poisson's equation and solving for $V_H$ (for instance, iteratively) and then injecting the obtained potential back into Eq. \eqref{eq:fstat}.

For the spin-Vlasov equations \eqref{eq:spin-vlasov1}-\eqref{eq:spin-vlasov2}, it is straightforward to see that, if we search stationary states for which all the spins are polarized in the same direction (say, the $z$ direction) and whose velocity distribution is isotropic, then the spin-Vlasov equations can be written as:
\begin{align}
 & \frac{\partial f_{0}}{\partial t} =
 \left\{  H^{\uparrow \uparrow} , f^{\uparrow \uparrow}  \right\} +  \left\{  H^{\downarrow \downarrow} , f^{\downarrow \downarrow}  \right\},
& \frac{\partial f_{z}}{\partial t} =  \left\{  H^{\uparrow \uparrow} , f^{\uparrow \uparrow}  \right\} - \left\{  H^{\downarrow \downarrow} , f^{\downarrow \downarrow}  \right\}.
\end{align}
where $\{ A,B\}$ denotes the Poisson brackets, $f_0=f^{\uparrow \uparrow}+f^{\downarrow \downarrow}$, $f_z=f^{\uparrow \uparrow}-f^{\downarrow \downarrow}$, and the Hamiltonians
$H^{\uparrow \uparrow}$ and $H^{\downarrow \downarrow}$ are defined as follows:
\begin{align}
H^{\uparrow \uparrow} = \frac{m}{2}|\bm{v}|^{2} - eV_H + \mu_{B}B_{z},~~~~
H^{\downarrow \downarrow} = \frac{m}{2}|\bm{v}|^{2} - eV_H - \mu_{B}B_{z}.
\end{align}
This implies that, at equilibrium ($\partial_t=0$), $f^{\uparrow \uparrow}$ should be a function of $H^{\uparrow \uparrow}$ and $f^{\downarrow \downarrow}$ a function of $H^{\downarrow \downarrow}$.
Therefore the stationary solutions are given by
\begin{align}
f_{0}^{stat} = \alpha^{\uparrow}  \left[ \mathcal{F_{D}} ( H^{\uparrow \uparrow} ) +  \mathcal{F_{D}} ( H^{\downarrow \downarrow} )\right]  ~~~\textrm{and} ~~~
f_{z}^{stat} = \alpha^{\downarrow} \left[ \mathcal{F_{D}} ( H^{\uparrow \uparrow} ) -  \mathcal{F_{D}} ( H^{\downarrow \downarrow} )\right].
\end{align}
The coefficients $\alpha^{\uparrow}$ and $\alpha^{\downarrow}$ are determined by imposing the correct normalization at $T=0$ for a non-interacting system. The free electron gas model predicts the following value for the spin-up and spin-down densities in the presence of an external and constant magnetic field \cite{Diu1989}:
\begin{align}
n^{\uparrow}(T=0)  &= \frac{1}{6\pi^{2}} \left(E_{F} - \mu_{B} B\right)^{3/2}, ~~~~
n^{\downarrow}(T=0) = \frac{1}{6\pi^{2}} \left(E_{F} + \mu_{B} B\right)^{3/2}.
\label{free electron t=0}
\end{align}
with $E_{F}$ the Fermi energy of the system, $E_{F} = \hbar ^{2}(3 \pi n_{0})^{2/3}/(2m)$.

In our case, the spin-up and spin-down densities are:
\begin{align}
n^{\uparrow}  &= \alpha^{\uparrow} \int \mathcal{F_{D}}\left(\frac{m}{2}\bm{v}^{2} + \mu_{B}B \right) d\bm{v} = 4 \pi \alpha^{\uparrow} \int_{0}^{\infty} \frac{v^{2} dv}{1+\exp \left[\left(\frac{m}{2}v^{2} + \mu_{B} B - \mu \right) / k_{B}T \right]}, \\
n^{\downarrow}&  = \alpha^{\downarrow} \int \mathcal{F_{D}}\left(\frac{m}{2}\bm{v}^{2} - \mu_{B}B \right) d\bm{v} = 4 \pi \alpha^{\downarrow} \int_{0}^{\infty} \frac{v^{2} dv}{1+\exp \left[\left(\frac{m}{2}v^{2} - \mu_{B} B - \mu \right) / k_{B}T \right]}.
\end{align}
In the limit $T \rightarrow 0$, one obtains
\begin{align}
n^{\uparrow}(T=0)  &= \frac{4 \pi}{3} \left( \frac{2}{3}\right)^{3/2}  \alpha^{\uparrow} \left( E_{F} - \mu_{B}B\right)^{3/2}, ~~~~~
n^{\downarrow}(T=0)  = \frac{4 \pi}{3} \left( \frac{2}{3}\right)^{3/2}  \alpha^{\uparrow} \left( E_{F} + \mu_{B}B\right)^{3/2}.
\label{free electron t=0 our}
\end{align}
Comparing Eqs. \eqref{free electron t=0} and \eqref{free electron t=0 our}, we deduce that
\begin{equation}
\alpha^{\uparrow} = \alpha^{\downarrow} = \left( \frac{m}{2\pi \hbar} \right)^{3}.
\end{equation}

We can also include exchange and correlations in the stationary state. For a collinear system (in the $z$ direction), one just needs to make the following transformations:
$V \mapsto V + V_{xc}$ and $B_{z} \mapsto B_{z} + (B_{xc})_{z}/\mu_{B}$.
In conclusion, the stationary states $f_{0}^{stat}$ and $f_{z}^{stat}$ can be written as:
\begin{eqnarray}
f_{0}^{stat} &=&\left(\frac{m}{2 \pi \hbar}\right)^{3} \Big{[} \mathcal{F_{D}} \left(  \frac{m}{2}|\bm{v}|^{2}  -eV_H + V_{xc} + \mu_{B} \left( B + B_{xc}\right)_{z}  \right)\nonumber  + \mathcal{F_{D}} \left(  \frac{m}{2}\bm{v}^{2}  -eV_H + V_{xc} - \mu_{B} \left( B + B_{xc}\right)_{z}  \right) \Big{]}, \label{complete expression stationary states f0}\\
f_{z}^{stat} &=&\left(\frac{m}{2 \pi \hbar}\right)^{3} \Big{[} \mathcal{F_{D}} \left(  \frac{m}{2}|\bm{v}|^{2}  -eV_H + V_{xc} + \mu_{B} \left( B + B_{xc}\right)_{z}  \right)\nonumber - \mathcal{F_{D}} \left(  \frac{m}{2}\bm{v}^{2}   -eV_H + V_{xc} - \mu_{B} \left( B + B_{xc}\right)_{z}  \right) \Big{]}.
\label{complete expression stationary states fz}
\end{eqnarray}
This is the general form of the electronic ground state. In practice, as we shall only deal with a 1D geometry appropriate to thin films, we should integrate the 3D Fermi-Dirac functions over the transverse velocities, to obtain:
\begin{align}
\mathcal{F_{D}}^{\pm}&= \frac{2 \pi k_{B} T}{m} \left(\frac{m}{2\pi \hbar}\right)^{3}  \ln\left[1 + \exp \left( -\frac{1}{k_{b}T}\left(\frac{m}{2}v^{2}  -eV_H + V_{xc} \pm \mu_{B} \left( B + B_{xc}\right)_{z}  - \mu(T) \right) \right) \right].
\label{fermi_dirac 1D}
\end{align}

   \begin{figure}[ht]
   \begin{center}
   \includegraphics[height=6cm]{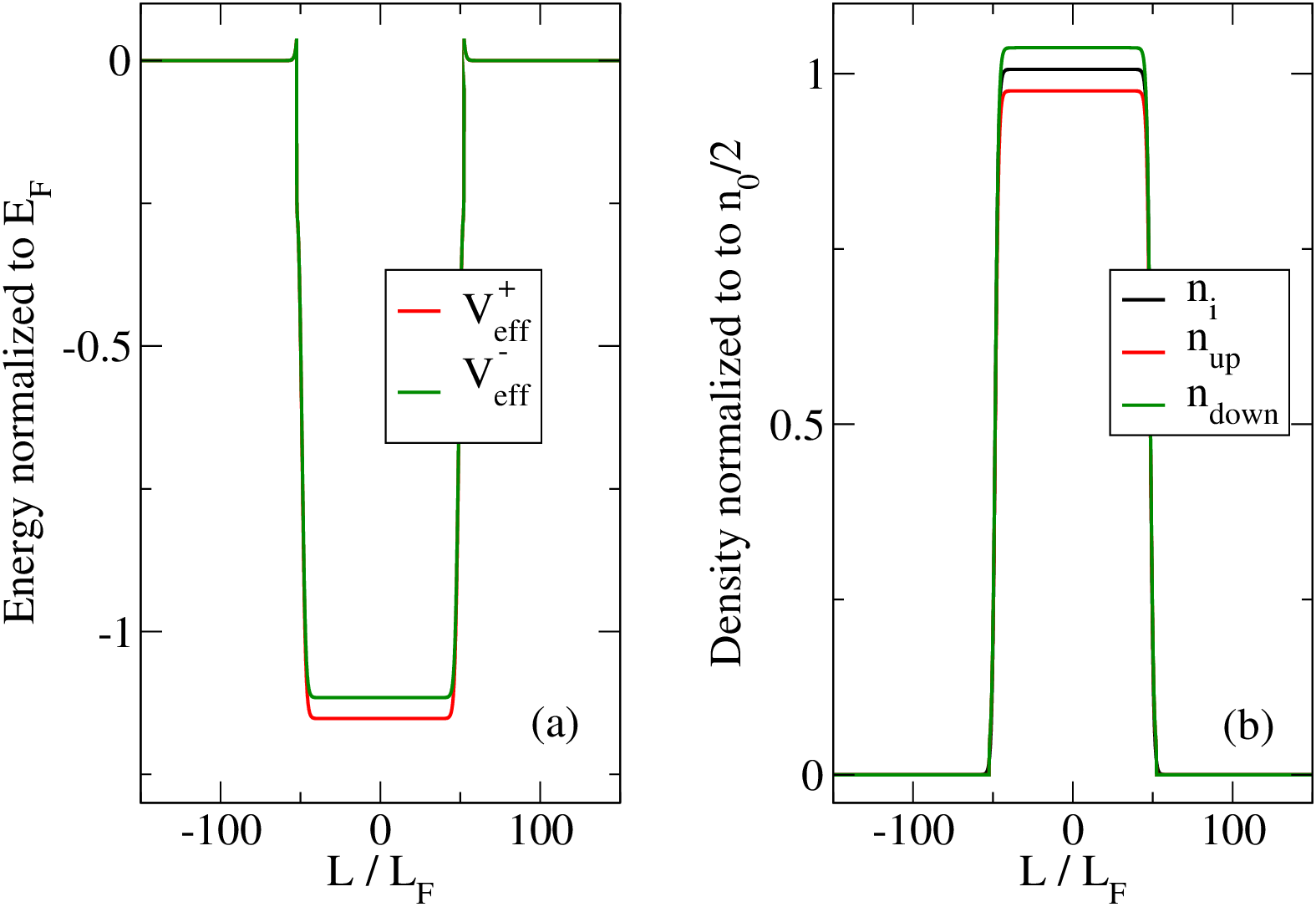}
   \end{center}
   \caption
   { \label{fig:ground} Left frame: Confinement potentials $V_{eff}^{\pm}$ for the spin-up (red curve) and spin-down (green curve) electrons. Right frame: Corresponding densities, together with the ion jellium density (black curve). The results are for a 8.4~nm film ($L=100L_F$) at $T=300\rm ~K$.
}
\vskip5mm
   \end{figure}

   \begin{figure}[ht]
   \begin{center}
   \includegraphics[height=6cm]{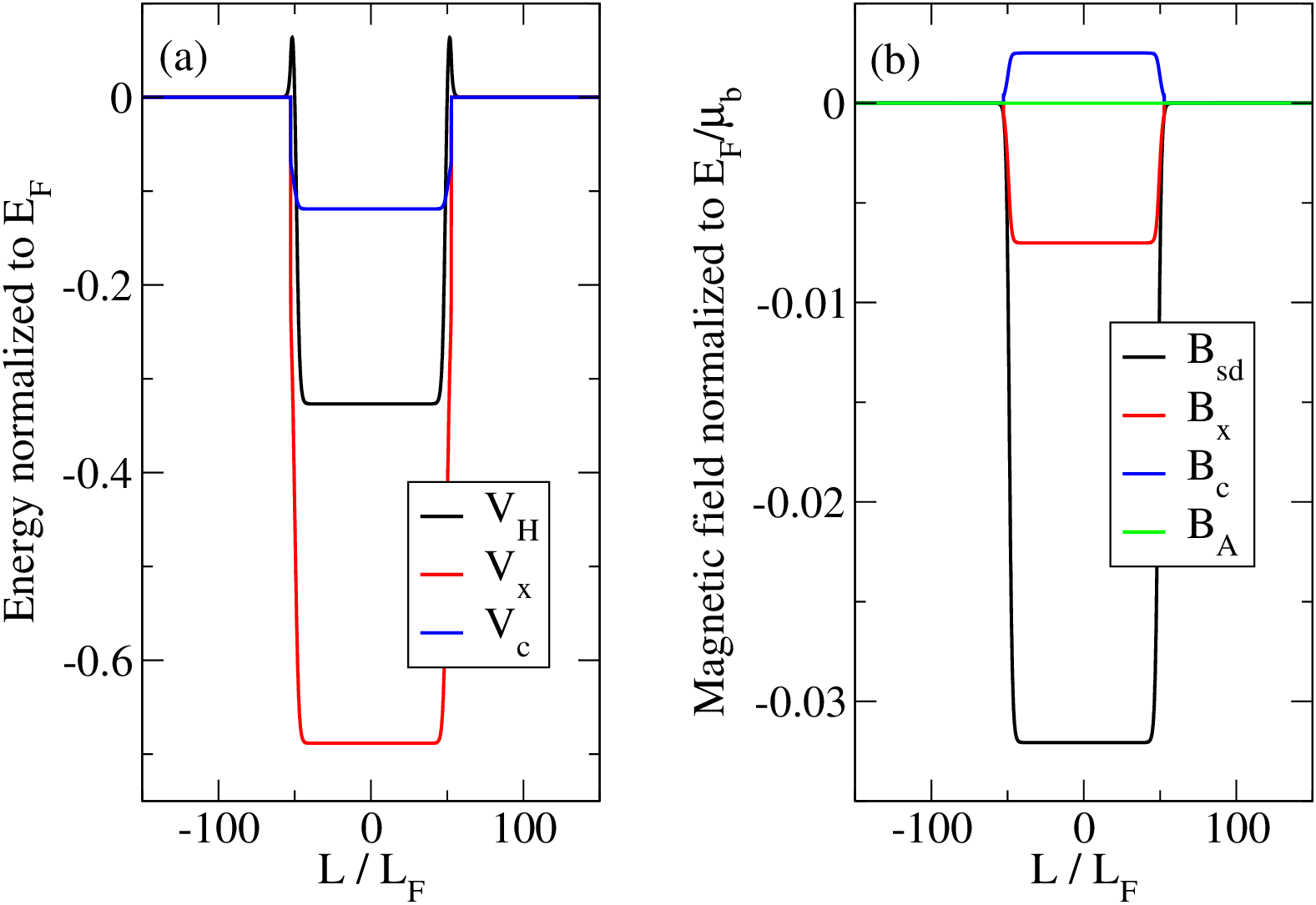}
   \end{center}
   \caption
   { \label{fig:ground2} Left frame: Electric potentials -- Hartree ($V_H$), electronic exchange $V_X$ and correlations $V_C$. Right frame: Magnetic field due to the electron-ion magnetic exchange $B_{sd}$, electronic exchange-correlations $B_{X, C}$, and self-consistent field arising from Ampère's equation $B_A$.
}
   \end{figure}

This is again a formal solution, as $V_H$ and $V_{xc}$ depend on the stationary distributions through the Poisson equation and the definition of $V_{xc}$ in terms of the electron density. An explicit solution is obtained using an iterative procedure.
Furthermore, when calculating the ground-state solution, the values of the magnetic exchange coupling constants $J$ and $K$ are fixed by imposing the correct Curie temperature for nickel ($T_{C}=631 \,\rm K$) and the correct ratio between the magnetization of the itinerant electrons ($\bm{M^e} = 0.066\,\mu_B/\rm atom$) \cite{Evans2014} and that of the localized ions ($\bm{M^i} = 0.54\,\mu_B/\rm atom$) \cite{Hurst2018}.

In Fig. \ref{fig:ground}, we show the computed ground-state densities and corresponding confining potentials, for the spin-up and spin-down components of the electron population in a 8.4~nm nickel film ($L=100L_F$) at room temperature.  Since the effective potentials $V_{eff}^{\pm}=-eV_H + V_{xc} \pm \mu_{B} \left( B + B_{xc}\right)_{z}$ differ for the two components, the respective densities also differ, and are slightly smaller or bigger than the positive ion density $n_i$.

\begin{figure}[ht]
   \begin{center}
   \includegraphics[height=5.5cm]{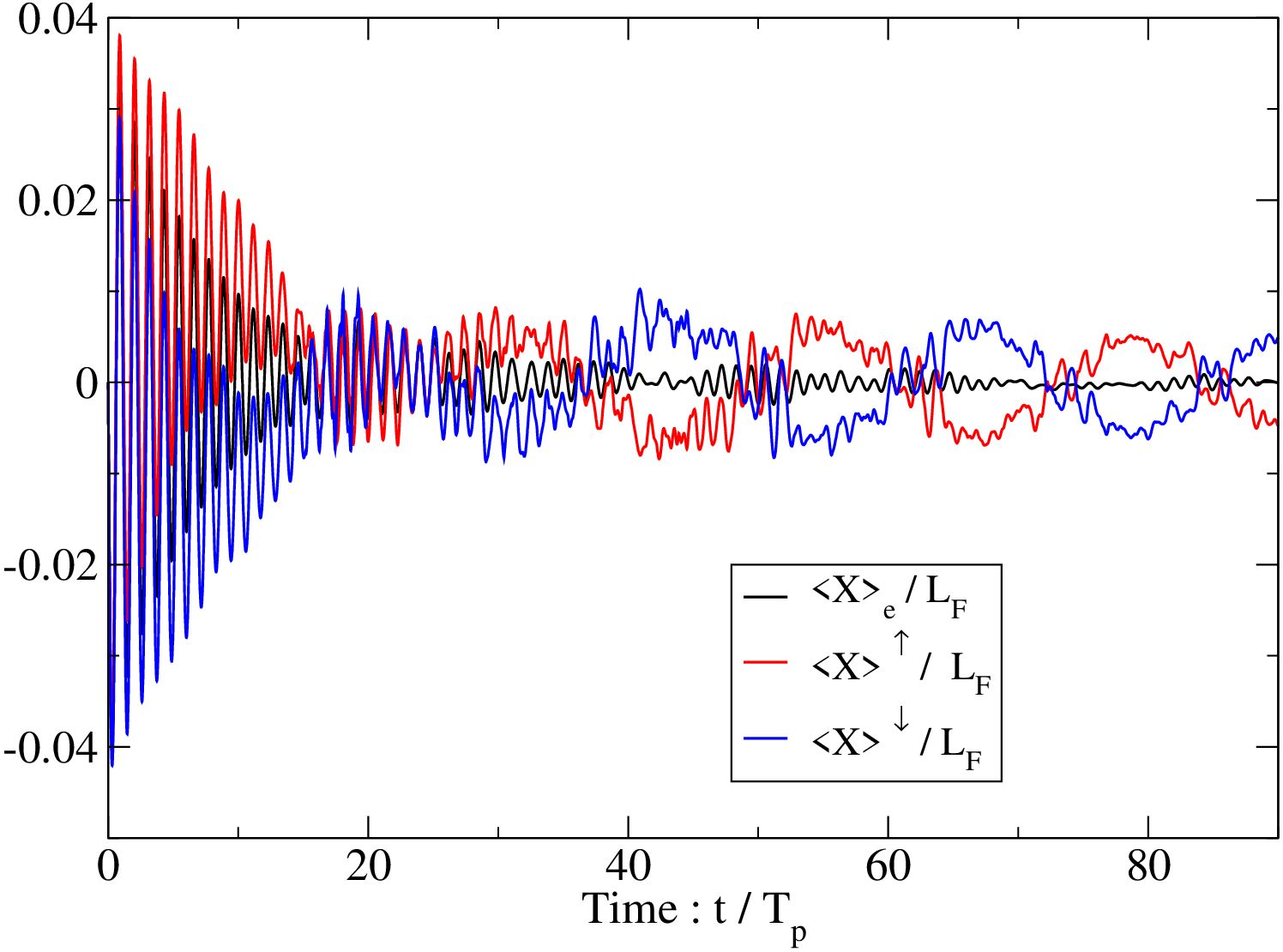}
   \hskip1cm \includegraphics[height=5.5cm]{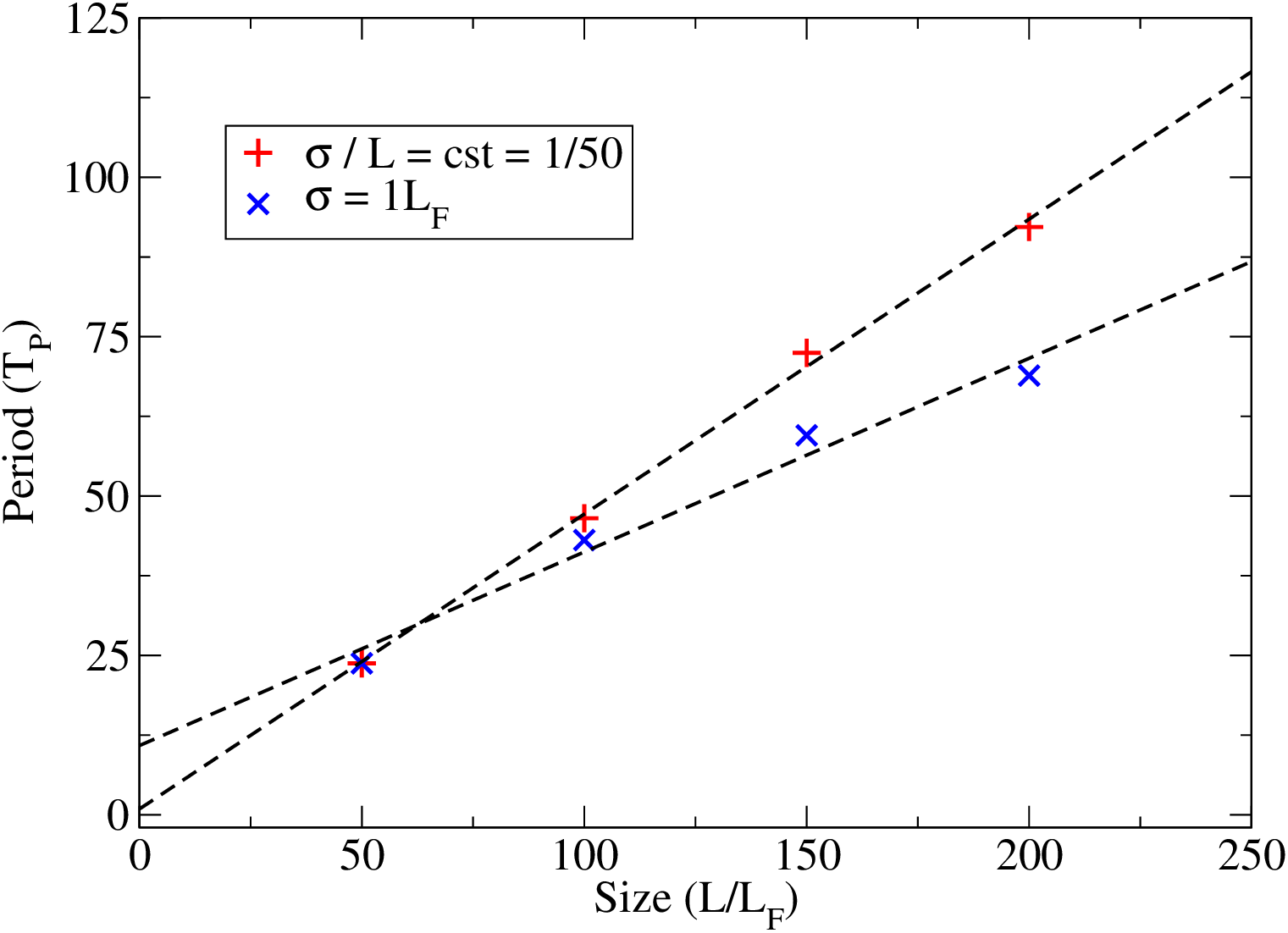}
   \end{center}
   \caption
   { \label{fig:dipoles} Left frame: Evolution of the electric dipole (black line) and the spin-up (red line) and spin-down (blue line) dipoles, as a function of time normalized to the plasmon period $T_p \approx \rm  0.26~fs$. The film thickness is $L=50L_F=4.2 \,\rm nm$.
   Right frame: Scaling of the magnetic dipole oscillation periods with the film thickness $L$, for a fixed soft-edge parameter $\sigma = L_F$ (blue crosses) and for a fixed ratio $\sigma/L$ (red plusses).}
\end{figure}

For the same film, Fig. \ref{fig:ground2} shows the various components of the electric potential (Hartree and electronic exchange-correlations) and magnetic field (the field due to the electron-ion magnetic exchange $B_{sd}$, as well the electronic exchange-correlations $B_{X, C}$, and the self-consistent magnetic field arising from Ampère's equation $B_A$). For the electric potential, the electron-electron exchange interaction is the dominant one, whereas for the magnetic properties the electron-ion magnetic exchange constitutes the largest contribution. The self-consistent Amp\`ere magnetic field is negligible.

\section{SPIN-CURRENT GENERATION}
\label{sec:spincurrent}

Let us first define the magnetic dipole: $\langle  X \rangle_{m} \equiv \int x f_{z} dx dv = \langle  X \rangle_{\uparrow}-\langle  X \rangle_{\downarrow}$, as well as the electric dipole: $\langle  X \rangle_{e} \equiv \int x f_{0} dx dv = \langle  X \rangle_{\uparrow}+\langle  X \rangle_{\downarrow}$. Just as the time derivative of the electric dipole corresponds to an electric current, the time derivative of the magnetic dipole represents a spin current.

As an illustrative example, we now show that an oscillating magnetic dipole can be generated by means of a purely electric excitation. The system is first prepared in its ground state, using the procedure described in Sec. \ref{sec:groundstate}. To excite the dynamics, we instantaneously accelerate all the electrons by a certain velocity shift $\delta v=0.05 v_F$, where $v_F$ is the Fermi speed of nickel (a more realistic excitation will be used later).

The main result is shown in Fig. \ref{fig:dipoles} (left frame), where we show the spin-up, spin-down, and electric dipoles as a function of time. The electric dipole decays away in a few tens of plasma periods mainly because of Landau damping; this effect had been noticed before on similar simulations of unmagnetized sodium films \cite{Manfredi2005film}. In contrast, the spin-up and spin-down dipoles continue to oscillate for much longer, virtually undamped. As the magnetic dipole is given by $\langle  X \rangle_{m} = \langle  X \rangle_{\uparrow}-\langle  X \rangle_{\downarrow}$, this implies that an oscillating magnetic dipole (and hence, a spin current) persists in the film for rather long times after any electric current has disappeared.

These magnetic dipole oscillations are due to ballistic electrons that travel in the film at a speed close to the Fermi velocity of nickel. Indeed, the period of the oscillations scales linearly with the size of the film, as is shown in the right panel of Fig. \ref{fig:dipoles}, for two cases where the soft-edge parameter $\sigma$ is kept fixed or varies proportionally to the film thickness. Similar ballistic charge oscillations had been previously observed in unmagnetized films \cite{Manfredi2005film}.

Next, we turn to a more realistic type of excitation, whereby the film is driven by short laser pulse. The electric field of the pulse is modeled as follows: $\bm{E}(x) = E_{0} \exp\left[-{(t-t_0)^2}/{2\Delta t^2} \right] \cos \left( \omega_L t\right) \widehat{\bm e}_{x}$,
where $E_{0}$ is the amplitude of the field, $\Delta t$ is its duration, $\omega_{L}$ the central frequency of the pulse, and $t_{0}$ is the time when it reaches its maximal amplitude.
For these simulations, we used a film of thickness $L=50L_F=4.2 \,\rm nm$ excited by a laser pulse of wavelength $\lambda_{L} = 2\pi c / \omega_{L} = 690 \,\rm nm$, and duration $2\Delta t= 10 \rm\, fs$, $t_{0}=15\rm \, fs$.
The laser period $2\pi/\omega_{L}=2.3 \,\rm fs$ is close to the ballistic period $2L/v_F =4.14 \,\rm fs$ (i.e., the time taken by the electrons to travel back and forth in the film at the Fermi speed), but far from the plasmon period $T_p = 0.26\,\rm fs$. In this way, the chances to principally excite the magnetic dipole oscillations are maximal \cite{Hurst2018}.

\begin{figure}[ht]
   \begin{center}
   \includegraphics[height=9cm]{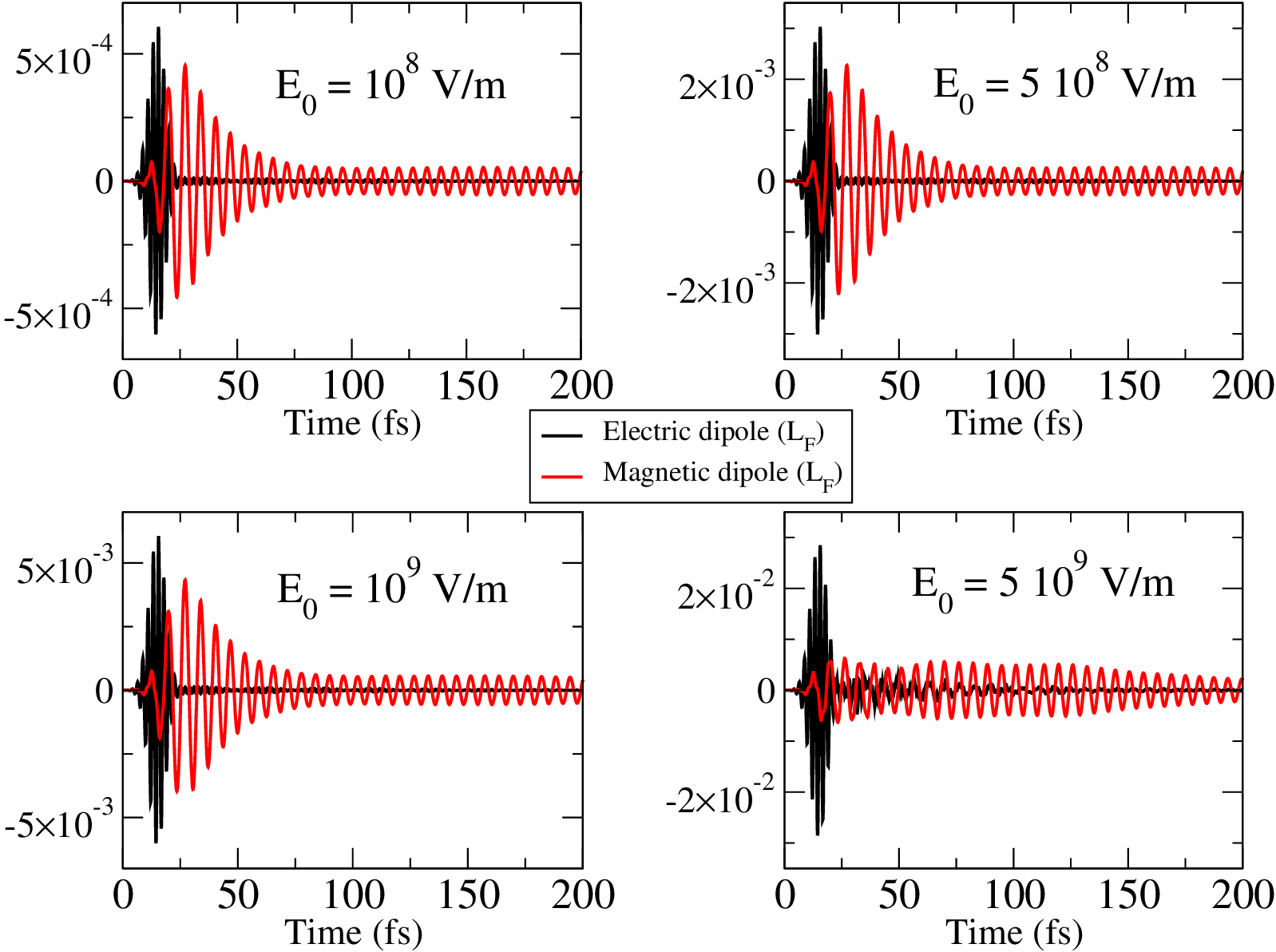}
   \end{center}
   \caption
   { \label{fig:E-scaling}
Electric (black curves) and magnetic (red curves) dipole evolutions for different values of the laser electric field amplitude $E_0$.
}
   \end{figure}

\begin{figure}[ht]
   \begin{center}
   \includegraphics[height=7cm]{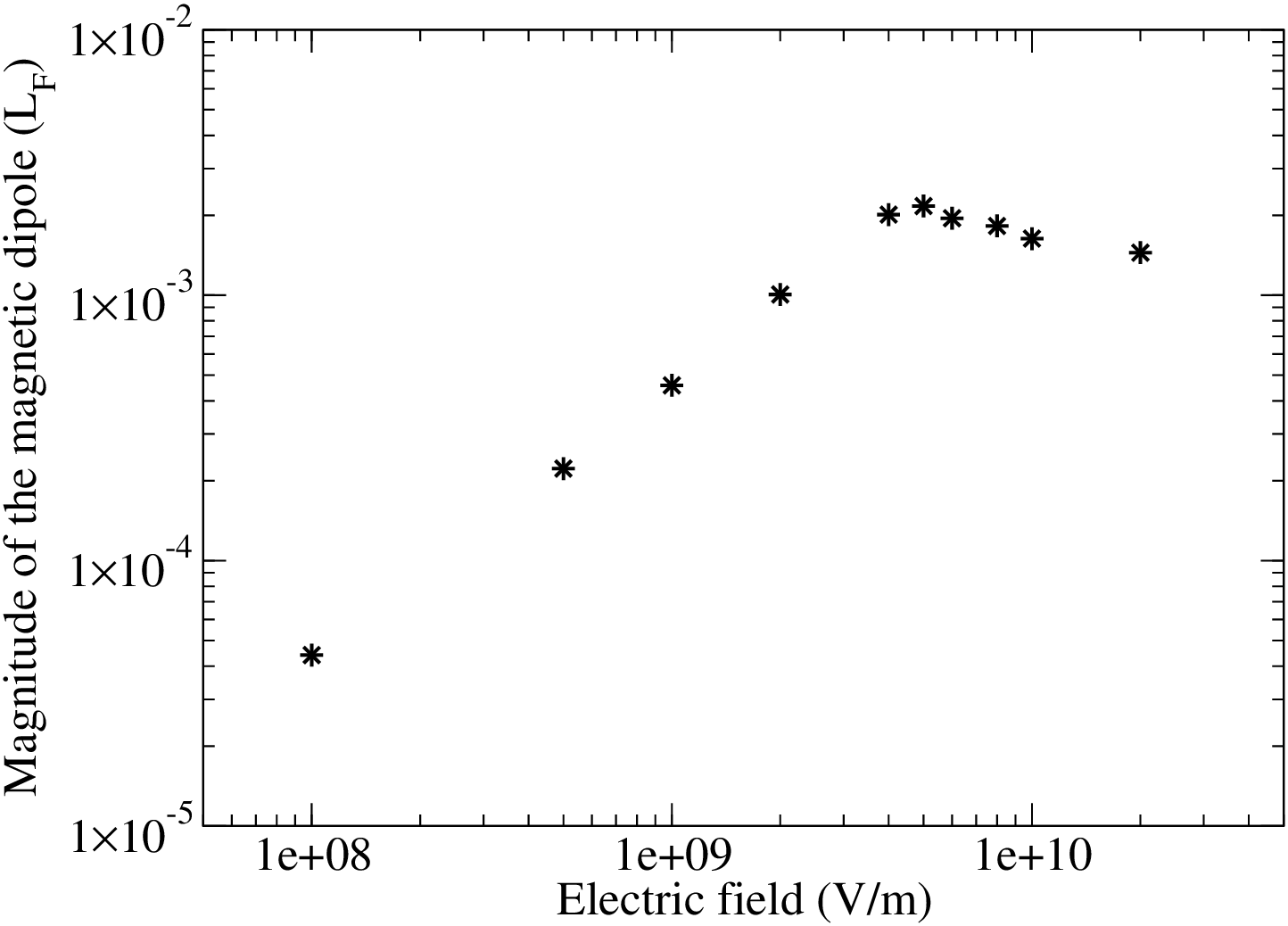}
   \end{center}
   \caption
   { \label{fig:dipole-scaling}
Amplitude of the magnetic dipole (averaged between $30\,\rm fs$ and $300\,\rm fs$) against the laser electric field $E_0$.
}
   \end{figure}

The amplitude of the laser electric field is varied in the range: $E_{0} = (10^8 - 10^{10})\rm \, V/m$, which covers both the linear and the nonlinear regime. In Fig. \ref{fig:E-scaling}, we show the electric and magnetic dipole evolutions for four different values of $E_0$. As in the preceding simulations, the electric dipole decays quickly after the laser has bee turned off. In contrast, the magnetic dipole persists for much longer times and its amplitude appears to increase with the amplitude of the electric field.
Figure \ref{fig:dipole-scaling} shows the amplitude of the magnetic dipole (averaged between 30 and 300 fs) against the electric field of the laser pulse. The dipole amplitude increases linearly with the field up to about $E_0=5\times 10^9\,\rm V/m$, after which it appears to saturate. The saturation arguably follows from the occurrence of nonlinear effects, but more studies would be needed to ascertain its origin with certainty.

\section{CONCLUSIONS}
\label{sec:conclusion}

In this work, we presented one of the first realistic studies of the charge and spin dynamics in ferromagnetic nano-objects. Our approach relied on the distinction between localized magnetism (carried by the fixed ions) and itinerant magnetism (carried by the delocalized electrons). The former was modeled by a LLG equation, whereas the itinerant magnetism, together with the charge dynamics, was modeled through a set of spin-Vlasov equations evolving the $(x,v)$ phase space. The model is semiclassical regarding the orbital motion, but fully retains the quantum-mechanical character of the spin degrees of freedom.

This approach is rather powerful, inasmuch as it allowed us to correctly describe the ferromagnetic ground state of nickel films without any free fitting parameters other than the magnetic exchange coupling constants (which are univocally determined by the magnetic properties of the system \cite{Hurst2018}).
Even more interestingly, it can be used to describe the coupled charge/spin dynamics in the time domain, including orbital magnetism, nonequilibrium effects, exchange and correlations (via the appropriate functionals), and quantum effects (through the Wigner matrix approach).

It would be straightforward -- albeit computationally costly -- to incorporate relativistic effects, such as the spin-orbit coupling, into the model. Further, non-conservative effects may be added in a relatively simple fashion by augmenting the spin-Vlasov equation with dissipative terms {\it \`a la} Fokker-Planck, as was done for nonmagnetic films \cite{Jasiak2010}. Applications to multilayer structures can also be envisaged without any conceptual difficulty.
All in all, our phase-space approach constitutes a valid and promising alternative to existing methods to study the ultrafast charge and spin dynamics in a variety of metallic nanostructures.

% \acknowledgments % equivalent to \section*{ACKNOWLEDGMENTS}

% References
\bibliography{biblio} % bibliography data in report.bib
\bibliographystyle{spiebib} % makes bibtex use spiebib.bst

\end{document}